\documentclass[aps,prl,twocolumn,superscriptaddress,preprintnumbers]{revtex4}

\usepackage{epsf,epsfig,amsmath,amssymb,amsfonts}
\usepackage{color}
\usepackage{slashed}

\newsavebox{\ns}
\newsavebox{\dbrane}

\def\be{\begin{equation}}
\def\ee{\end{equation}}
\def\bea{\begin{eqnarray}}
\def\eea{\end{eqnarray}}

\def\Dslash{\,\,{\raise.15ex\hbox{/}\mkern-12mu D}}
\def\Dbarslash{\,\,{\raise.15ex\hbox{/}\mkern-12mu {\bar D}}}
\def\delslash{\,\,{\raise.15ex\hbox{/}\mkern-9mu \partial}}
\def\delbarslash{\,\,{\raise.15ex\hbox{/}\mkern-9mu {\bar\partial}}}
\def\pslash{\,\,{\raise.15ex\hbox{/}\mkern-9mu p}}
\def\calDslash{\,\,{\raise.15ex\hbox{/}\mkern-12mu {\cal D}}}

\newcommand{\vol}{\mbox{vol}}

\newcommand{\nn}{\nonumber \\}

\def\a{\alpha}

\def\G{\Gamma}

\def\G{\Gamma}

\def\e{\epsilon}
\def\s{\sigma}
\def\e{\epsilon}

\allowdisplaybreaks

\begin{document}

\title{Supersymmetric $AdS_6$ via T-duality}

\preprint{DMUS-MP-12/10}         

\preprint{FPAUO-12/19}   

\vskip 1 cm

 \author{Y. Lozano}
 \affiliation{Departamento de F\'isica, Universidad de Oviedo, Oviedo 33007, Spain}
 \author{E. \'O Colg\'ain}
 \affiliation{Departamento de F\'isica, Universidad de Oviedo, Oviedo 33007, Spain}
 \author{D. Rodr\'iguez-G\'omez}
 \affiliation{Departamento de F\'isica, Universidad de Oviedo, Oviedo 33007, Spain}
\author{K. Sfetsos}
\affiliation{Department of Mathematics, University of Surrey, Guildford GU2 7XH, UK}
\affiliation{Department of Engineering Sciences, University of Patras, 26110 Patras, Greece}

\begin{abstract}
We present a new supersymmetric $AdS_6$ solution of type IIB supergravity with $SU(2)$ isometry. Through the $AdS/CFT$ correspondence, this has potentially very interesting implications for 5d fixed point theories. This solution is the result of a non-Abelian T-duality on the known supersymmetric $AdS_6$ solution of massive IIA. The $SU(2)$ R-symmetry is untouched, leading to sixteen supercharges and preserved supersymmetry.
\end{abstract}

\maketitle

\setcounter{equation}{0}

\section{Introduction} \label{Introduction}

Gauge theories in five dimensions (5d) are, at least naively, non-renormalizable and therefore uninteresting as microscopic theories. However, on the contrary, despite this gloomy conclusion 5d gauge theories lead to very interesting phenomena. 
In particular, quite remarkably, it is possible to find 5d gauge theories as consistent theories \textit{per se} \cite{Seiberg:1996bd,Morrison:1996xf,Intriligator:1997pq}. These 5d fixed point theories are intrinsically strongly coupled and can exhibit very exotic phenomena such as exceptional global symmetry groups arising from non-perturbative effects. These theories not only play a very important role in understanding crucial aspects of string theory --\textit{e.g.} \cite{Polchinski:1995df}--, but also in helping us understand the landscape of quantum field theories in general.

Being intrinsically strongly coupled, 5d fixed point theories are hard to study through more traditional methods. On the other hand, in view of the success of the $AdS_{d+1}/CFT_d$ duality in unravelling mysteries about 3d and 4d superconformal field theories it is natural to apply the latest holographic methods to the 5d case. In fact, this case has been largely overlooked to date, even though very recently  there has been a steady stream of recent developments along these lines \cite{Bergman:2012kr,Jafferis:2012iv,Kim:2012gu,Bergman:2012qh}. In particular, through the $AdS_6/CFT_5$ duality, we might search for new  $CFT_5$ by scanning over the possible $AdS_6$ vacua in supergravity. Remarkably, up to now only one supersymmetric $AdS_6$ solution  \cite{Brandhuber:1999np} was known; the existence of which was anticipated by \cite{Ferrara:1998gv}, and orbifolds therereof \cite{Bergman:2012kr}. Indeed, it has also recently been confirmed that this supersymmetric solution (and its orbifolds) is unique in massive type-IIA \cite{Passias:2012vp}. Following this philosophy, in this note we exhibit a new $AdS_6$ solution in type-IIB. Even though a full understanding of its features is still lacking, through the holographic correspondence it is natural to expect this new solution to be very relevant for defining a new class of 5d fixed point theories.

Our new $AdS_6$ solution is produced by performing a non-Abelian T-duality transformation on the known $AdS_6$ solution of  \cite{Brandhuber:1999np}. Given a non-linear sigma-model (NLSM) with a target space-time geometry admitting an Abelian isometry, a well-defined prescription exists 
for gauging the isometry, integrating out the gauge field and producing the so-called T-dual sigma-model  \cite{Buscher:1987qj,Rocek:1991ps}. 
Then, from the T-dual sigma-model it is possible to infer how the geometry changes under this T-duality transformation. 
The beauty of the gauging approach is that it is immediately generalisable beyond the Abelian case to both  non-Abelian isometries, 
early accounts of which appear in \cite{delaOssa:1992vc,Giveon:1993ai,Sfetsos:1994vz,Alvarez:1994np},
 and more recently, fermionic isometries \cite{fermTdual, Beisert} (see \cite{OColgain:2012si} for a recent review).

Non-Abelian T-duality has only recently been upgraded to a symmetry of type II supergravity \cite{Sfetsos:2010uq,Lozano:2011kb}, so new supergravity solutions can be generated from old ones. 
In contrast to Abelian T-duality, non-Abelian T-duality may not be regarded as a symmetry of string theory, and it has notable quirks.
For instance, it is not clear how to constrain holonomies of gauge fields and show that the original and T-dual models have the same path integrals.
However, a user-friendly description of the $SU(2)$ transformation \cite{Itsios:2012dc} allows one to plug in a space-time with an $SO(4)$ isometry and generate a T-dual solution. 
In the process the chirality of the theory flips, i.e. from type-IIA to type-IIB and vice versa {\cite{Sfetsos:2010uq,Lozano:2011kb}}.
Ref. {\cite{Itsios:2012dc}} also shows that we can understand non-Abelian T-duality in terms of inert lower-dimensional theories that are invariant 
under the duality, just as in the Abelian case \cite{Bergshoeff:1995as}.

%

As described, the main object of this note is to draw attention to another supersymmetric solution that can be constructed from the literature. While it is expected that Abelian T-duality on the $AdS_6 \times S^4$ solution of massive IIA produces a supersymmetric solution of type IIB with $SU(2) \times U(1)$ isometry, here we show that, following \cite{Itsios:2012dc}, by performing an $SU(2)$ non-Abelian T-duality the resulting background is a new supersymmetric solution to type IIB with just $SU(2)$ isometry. This observation has profound potential implications for the existence of new fixed point theories in 5d with a gravitational dual. Moreover, our solution is novel in a further regard; it is the first example of a non-Abelian T-dual geometry with supersymmetry fully preserved. 

\newpage

\section{D4-D8 near-horizon}
The only known supersymmetric $AdS_6 \times S^4$ solution of massive IIA supergravity \cite{Romans:1985tz} arises as the near-horizon of D4-D8 \cite{Brandhuber:1999np}.

The string frame solution is
\bea
\label{warp}
ds^2 &=& \frac{1}{4} W^2 L^2 \left[ 9 ds^2(AdS_6) + 4 d s^2(S^4) \right] \ , \nn
F_4 &=& 5 L^{4} (m \cos \theta)^{1/3} \sin^3 \theta d \theta \wedge \vol(S^3)\ , \nn
e^{\Phi} &=& \frac{2 }{3 L (m \cos \theta)^{5/6}}\  ,
\eea
where $m$ is the Romans' mass, $L$ denotes the $AdS_6$ radius, $W$, the warp factor, is a function of $\theta$, $W =  (m \cos \theta)^{-1/6}$,
and the metric on $S^4$ takes the form
\bea
ds^2(S^4) &=&  d \theta^2 + \sin^2 \theta ds^2(S^3)\ .
\eea
While $S^4$ would have $SO(5)$ isometry, the $\theta$-dependent warping means that this is broken to $SO(4) \sim SU(2)_{G} \times SU(2)_{R}$, where one $SU(2)_{G}$ is 
a global symmetry and the other an R-symmetry. In addition, as the range for $\theta$ is $0 \leqslant \theta \leqslant \pi/2$, instead of a whole $S^4$, we only have half, 
and at one end-point of this range, $\theta = \pi/2$, the warp factor $W$ blows up leading to a curvature singularity.
{In addition the string coupling $e^\Phi$ blows up.}

\section{Non-Abelian T-duality}
The non-Abelian dual of a general class of type-II supergravity solutions with isometry $SO(4) \sim SU(2) \times SU(2)$, with respect to any of these $SU(2)$ 
subgroups, was given in
\cite{Itsios:2012dc}. Moreover, it was demonstrated that the original solution on $S^3$, and the T-dual solution on the dual space $M_1 \times S^2$ (see below), 
may be reduced consistently to give the \textit{same} theory in seven-dimensions \cite{Itsios:2012dc}, thus offering another perspective on the fact  that non-Abelian T-duality is a symmetry of the equations of motion.

Recall from \cite{Itsios:2012dc} that given a massive type-IIA solution of the form
\bea
ds^2_{IIA} &=& ds^2(M_7) + e^{2A} ds^2(S^3)\ ,
\nn
F_0 &=& m\ ,
\nn
F_2 &=& G_2\ , \\
F_4 &=& G_1 \wedge \vol(S^3) + G_4\ ,
\nonumber
\eea
where $m$ is the mass, $A$ is a scalar warp factor and the $B$-field, dilaton, $\Phi$, and the $n$-form fluxes, $G_n$, just depend on the seven-dimensional space-time, 
the NS sector of the type-IIB supergravity $SU(2)$ T-dual is given by
\bea
\label{nonabdual}
d\hat{s}^2_{IIB} &=& ds^2(M_7) + e^{-2A} dr^2 + \frac{r^2 e^{2A}}{r^2 + e^{4A}} ds^2(S^2)\ ,
\nn
\hat{B} &=&  B + \frac{r^3}{r^2 + e^{4A}} \vol (S^2)\ ,
\\
e^{-2 \hat{\Phi}} &=& e^{-2 \Phi} e^{2A} (r^2 + e^{4A})\ ,
\nonumber
\eea
where we have introduced hats to differentiate T-dual fields from those of the original solution. Observe that in the process of doing the $SU(2)$ transformation, 
one of the $SU(2)$ isometries is selected out and gets broken, leaving a manifest residual $SU(2)$ isometry in the form of the remaining two-sphere.  
In turn, (\ref{nonabdual}) is a solution of the type-IIB equations of motion for any positive value of $r$. In order to fully clarify the nature of the space 
spanned by this variable we should resort to the sigma-model derivation of T-duality. However it is not clear how to extract global topological properties in the non-Abelian case \cite{Alvarez:1993qi}.
The complementing general expressions for the RR fluxes post T-duality may be found in \cite{Itsios:2012dc}, and owing to their length, we omit them.

Although the equations of motion are guaranteed to be satisfied, more pertinent to our current discussion is the issue of preserved supersymmetry.
 From \cite{Itsios:2012dc} we know that under an $SU(2)$ transformation from type IIB supergravity to massive IIA the Killing spinor equations may 
be mapped up to the gravitino variation in the $r$-direction. Interestingly, this single expression encapsulates all the information about the projection 
conditions on the Killing spinors of supersymmetry preserving T-duals. It is certainly expected that for transformations from massive IIA to type-IIB the supersymmetry 
conditions also simply boil down to one condition. Indeed, some work reveals this is the case and through the usual rotation of the type-IIB Killing spinor
\be
\label{rotate}
\eta =  e^{X} \tilde{\eta} =  \exp \left( - \frac{1}{2} \tan^{-1} \left( \frac{e^{2A}}{r} \right)
\Gamma^{\alpha_1 \alpha_2} \sigma^3 \right) \tilde{\eta}\ ,
\ee
where $\alpha_i$, $i=1,2$ denote coordinates on the residual $S^2$, one can demonstrate that if the original geometry is supersymmetric, 
then the T-dual geometry is also supersymmetric provided
\bea
\label{psir}
\delta \psi_{r} &=& e^{X} \biggl[ \frac{1}{2} \slashed{\partial} A \Gamma_{r} - \frac{e^{-A}}{4} \Gamma^{\alpha_1 \alpha_2} \sigma^3 + \frac{e^{\phi}}{8} 
\biggl(  m i \sigma^2  \nn &+& e^{-3 A} \slashed{G}_1 \G^{r \a_1 \a_2} \s^1  + \slashed{G}_2 \s^1  - \slashed{G}_3 \G^{r \a_1 \a_2} i \s^2 \biggr) \biggr] \tilde{\eta}
 \nn
&=& 0\ , 
\eea
where we have defined $G_3 = *_7 G_4$. Note that $\eta$ is further decomposed in terms of real Majorana-Weyl spinors
\be
\eta = \left( \begin{array}{c} \e_+ \\ \e_- \end{array} \right)\ .
\ee

\section{T-dual $AdS_6$}
Before performing an $SU(2)$ transformation on (\ref{warp}), we comment on the $U(1)$ T-duality in the same context. A natural $U(1)$ direction can be found by rewriting the metric on $S^3$ in terms of a Hopf-fibre over $S^2$
\be
ds^2 = \frac{1}{4} \left[ d \phi_1^2 + \sin^2 \phi_1 d \phi_2^2 + (d \phi_3 + \cos \phi_1 d \phi_2)^2 \right]\ .
\ee
Here $\phi_3$ labels the Hopf-fibre direction, T-duality on which has previously been discussed in the literature in \cite{Cvetic:2000cj}, 
without commenting on the preserved supersymmetry. Indeed, the Killing spinors for the original $AdS_6 \times S^4$ solution with this parameterisation of the $S^3$ take the form
\be
\eta = (\cos \theta)^{-1/12}  e^{-\frac{\theta}{2} \gamma \G^{\theta} \s^1} e^{-\frac{\phi_1}{2} \G^{\phi_3 \phi_2} } e^{-\frac{\phi_2}{2} \G^{\phi_2 \phi_1} } \tilde{\eta}\ ,
\ee
where  $\gamma = \G^{\theta \phi_1 \phi_2 \phi_3}$ and $\tilde{\eta}$ denotes the Killing spinor on $AdS_6$. The Killing spinor is subject to a single projection condition
\be
\label{ads6s4proj1}
\left[ \sin \theta \G^{\theta} \s^1 + \cos \theta \G^{\theta \phi_1 \phi_2 \phi_3}  \right] \eta = - \eta\ ,
\ee
so we have sixteen supersymmetries, the minimum required for a supersymmetric $AdS_6$ geometry. Furthermore, as is evident from the explicit 
form of the Killing spinor, it is independent of $\phi_3$, so that when one performs the Abelian T-duality one expects no supersymmetry to be broken. 
By explicitly working out the Killing spinor equations for the Abelian T-dual one can also confirm this to be the case.  So supersymmetric $AdS_6$ geometries in type-IIB certainly exist.

The main result of this letter now follows. The $U(1)$ Hopf-fibre T-duality produces a supersymmetric T-dual because we are simply picking out 
a $U(1)$ direction from the $SU(2)$ global symmetry. Therefore, in the process of doing the T-duality, the $SU(2)$ R-symmetry is untouched. 
Now, we also have the freedom to do an $SU(2)$ T-duality using the full global symmetry. Again the rational is the same; as we do not touch the R-symmetry we are guaranteed to produce a supersymmetric solution. So cranking the handle, one arrives at
\bea
\label{warpsu2}
d \hat{s}^2 &=&  \frac{1}{4} W^2 L^2 \left[ 9 ds^2(AdS_6) + 4 d \theta^2 \right] \nn &&+ e^{-2 A} dr^2 + \frac{r^2 e^{2 A} }{ r^2 + e^{4 A}} ds^2(S^2)\ ,
\nonumber\\
 \hat B &=&  \frac{r^3}{r^2 + e^{4 A}} \vol(S^2)\ ,
 \nn
 e^{-2 \hat \Phi} &=& e^{-2 \Phi} e^{2 A} (r^2 + e^{4 A})\ ,
\nonumber \\
\hat F_1 &=& -G_1 - m r dr\ ,
\\
\hat F_3 &=& \left[ - \frac{r^3}{r^2 + e^{4 A}} G_1  + \frac{mr^2 e^{4A}}{r^2 + e^{4A}} dr \right] \wedge \vol(S^2)\ ,
\nonumber
\eea
where we have introduced the following
\be
e^A = \frac{W L \sin \theta}{2}, \quad G_1 = \frac{5}{8} L^4 (m \cos \theta)^{1/3} \sin^3 \theta d \theta\ .
\ee
At $\theta=0$, just as with the Abelian T-dual,  there is a curvature singularity and $e^{\hat \Phi}$ blows up. This is in addition to the
singularity at $\theta=\pi/2$ inherited from the original solution.

We are now in a position to plug this solution back into (\ref{psir}), the only independent Killing spinor equation post T-duality, 
to identify the projection conditions on the Killing spinor. In the process, one encounters a single projection condition
\be
\label{su2proj}
\left[ \cos \theta \G^{\theta r \a_1 \a_2} \s^3 - \sin \theta \G^{\theta r} i \s^2 \right] \tilde{\eta} = - \tilde{\eta}\ ,
\ee
thus showing that supersymmetry is preserved. Moreover, by employing the redefinitions
\be
\label{redefine}
\tilde{\epsilon}_+ = \G^{r} \e_+, \quad \tilde{\e}_- = \e_-, \quad \G^{r \a_1 \a_2} = - \G^{\phi_1 \phi_2 \phi_3}\ ,
\ee
one can recover the original projector (\ref{ads6s4proj1}).

\section{Discussion}
While it can be rationalised at some level, i.e. we are not touching the R-symmetry, this indeed is a striking result. 
To appreciate this, recall that even for flat space-time, the $SU(2)$ T-duality transformation we have employed here breaks
 supersymmetry by one half \cite{Itsios:2012dc}. So, in the original warped supersymmetric $AdS_6 \times S^4$ solution of 
massive type-IIA, we have found the first example of a non-Abelian T-duality transformation where supersymmetry is preserved. 
It turns out that the example presented in this paper is however not unique. Other examples based on the Klebanov--Witten 
and Klebanov--Strassler ${\cal N}=1$ backgrounds, for which supersymmetry is also preserved under non-Abelian T-duality, 
have also been constructed in \cite{work2, Itsios:2013wd}. 

A pressing question concerns the AdS/CFT interpretation. The identification of the dual SCFT for the non-Abelian T-duality transformation is a long-standing problem and the 
jury is certainly out on whether one exists, and if it does, whether it is the same SCFT, or indeed a different theory. 
In the process of doing the $SU(2)$ T-duality in the $AdS_6 \times S^4$ context, the $SU(2)$ global symmetry is completely broken,
 leaving just the R-symmetry. For the Abelian T-duality the isometry is also reduced, but there we are confident that the theory does not change. In fact, the Cartan  of the isometry group remains the same. On the other hand, in the $SU(2)$ T-dual such Cartan subgroup seems different than that of the original background suggesting that the dual theory --if it exists-- would be different.
 Moreover, as opposed to the standard Abelian T-duality transformation, in this case the size of the internal space $M_1$ appears as an extra parameter of the solution. While the implications of this striking new feature remain to be uncovered, this seems to imply that the dual CFT should contain one extra charge as compared to the original one. One possible way out would be to take the size of $M_1$ to infinity.  However, this would lead most likely to a puzzling feature in the dual CFT, namely, a continuous spectrum. Although further checks are certainly required to elucidate these and other properties of the dual CFT (progress is underway in [30]) it is clear that the dual CFT faces new challenges whose resolution will help understanding the role played by non-Abelian T-duality in the context of the AdS/CFT correspondence (see also \cite{work2, Itsios:2013wd} for the study of the SCFTs associated to 
 the non-Abelian duals of the Klebanov--Witten and Klebanov--Strassler backgrounds). 
 
Moreover, now that we have two distinct solutions in type-IIB, it may be an opportune time to build on the work initiated
 in massive IIA \cite{Passias:2012vp} and classify the supersymmetric solutions in this setting also. A priori we will have at least two branches, one with $U(1)$ T-dual 
 and the other with the $SU(2)$ T-dual.

Finally, another interesting direction for study concerns the KK reduction \cite{Cvetic:1999un} from massive IIA on $S^4$ 
to Romans' F(4) supergravity \cite{Romans:1985tw}. In \cite{Itsios:2012dc} it was shown that there was a consistent truncation to $D=7$. 
The only terms of the KK reduction inconsistent with $SU(2)$ T-duality as described in \cite{Itsios:2012dc} are the $SU(2)$ gauge fields. 
So, as it stands, any solution to Romans' theory now also uplifts to a solution to type-IIB provided the $SU(2)$ gauge fields are not excited. 
In this sense, here we are simply discussing the supersymmetric $AdS_6$ vacuum. We can think of putting the gauge fields back in 
if we gauge the residual $SU(2)$ R-symmetry of the non-Abelian T-dual. This all echoes well with the conjecture \cite{Gauntlett:2007ma}  that gauging the R-symmetry always leads to a consistent reduction.


\section*{Acknowledgements}	
We have enjoyed conversations with A. Passias and D.C. Thompson. D. R-G is grateful to the Technion Department of Physics for hospitality while this work was being completed, and especially to Oren Bergman. Y. L, E. \'O C and D. R-G are partially supported by the research grants MICINN-09-FPA2009-07122 and MEC-DGI-CSD2007-00042. The work of D. R-G is partly supported by the Ramon y Cajal fellowship RyC-20011-07593 and the Ally Kaufmann fellowship at Technion (Israel). This research is implemented (K.S.) under the “ARISTEIA” action
of the “operational programme education and lifelong learning” and is co-funded by
the European Social Fund (ESF) and National Resources.

\end{document}